\pdfoutput=1
\documentclass{JINST}
\usepackage{cite}
\usepackage{booktabs}
\usepackage{url}
\usepackage{float}

\usepackage{enumitem}

\title{Performance of cryogenic charge readout electronics with the ARGONTUBE LAr TPC}

\author{ A.~Ereditato, D.~Goeldi, S.~Janos, I.~Kreslo\thanks{Corresponding author.}~, 
M.~Luethi, C.~Rudolf~von~Rohr, M.~Schenk, T.~Strauss, M.~S.~Weber and M.~Zeller\\
\llap Albert Einstein Center for Fundamental Physics, Laboratory for High Energy Physics\\
University of Bern, Switzerland\\
E-mail: \email{igor.kreslo@lhep.unibe.ch}}

\abstract{ARGONTUBE is a liquid argon time projection chamber (TPC) with an electron drift length of up to 5~m equipped with cryogenic charge-sensitive preamplifiers. In this work, we present results on its performance, including a comparison of the new cryogenic charge-sensitive preamplifiers with the previously used room-temperature-operated charge preamplifiers.}


\begin{document}
\section{Introduction}
The ARGONTUBE detector at the Laboratory for High Energy Physics (LHEP) in Bern is a liquid argon time projection chamber (LAr TPC), which allowed to achieve a drift distance for ionization electrons in liquid argon as long as 5~m~\cite{ARGONTUBE0,ARGONTUBE1,ARGONTUBE2} for the first time.
This detector was conceived as a demonstrator and test setup to address key features of future large-size LAr TPCs, such as life time of the drifting electrons, diffusion properties of the charge carriers, supply of high voltage to the cathode, as well as the identification and correction of possible drift field distortions by means of UV laser beams~\cite{Badhrees:2010zz}.

The typical amount of charge per readout channel resulting from the track of an ionizing particle in LAr TPCs is of the order of femto-Coulombs. Achieving a detector output signal with a reasonable signal-to-noise ratio for efficient track reconstruction in presence of different types of electronic noise requires the application of high-end low-noise front-end amplifiers. One important noise source in such systems is Johnson noise, which is temperature-dependent (see~\cite{Veljko} for a detailed analysis). Lowering the temperature of the amplification circuit leads to a significant reduction of this sort of noise. MOSFET transistors are known to perform well at temperatures down to several tens of Kelvin, and their parameters, such as leakage currents and transconductance, even improve with cooling (see $e.g.$ \cite{MOSFET1}). Interconnections between the charge readout electrodes and the preamplifier introduce an additional input capacitance that contributes to the noise. Therefore, the length of these connections must be minimized. As a result, the installation of cryogenic integrated circuits based on CMOS transistors as a first stage of charge amplification, placed as close as possible to the charge readout electrodes of the TPC and operating at liquid argon temperature (87~K), is a very attractive solution to improve the signal-to-noise ratio of the TPC output signal. 

In this paper, we present results obtained by running the ARGONTUBE detector, registering cosmic ray muons and UV laser beam tracks with a cryogenic charge readout system based on a dedicated application-specific integrated circuit (ASIC) developed by colleagues from the Brookhaven National Laboratory \cite{DGeronimo}\footnote{Brookhaven National Laboratory, Upton, NY 11973-5000.}. 

\section{ARGONTUBE and its charge readout systems}

For a detailed description of the ARGONTUBE detector and of its various subsystems, we refer to our previous publications~\cite{ARGONTUBE0,ARGONTUBE1,ARGONTUBE2}. The charge readout system described earlier in these papers was based on G5601 hybrid preamplifier boards, custom-designed at ETH Zurich\footnote{ETH Z\"{u}rich, R\"{a}mistrasse 101, 8092 Z\"{u}rich, Switzerland.} \cite{Badertscher}. These boards, in total 64 of them, each containing 2 channels, were modified to accommodate the larger bandwidth required for ARGONTUBE, and placed at the top flange of the cryostat and were operated at room temperature. A signal-to-noise ratio of about 3 was obtained for the signal induced by minimum ionizing particles (MIP) crossing the active volume in the proximity of the readout planes.

In order to reduce the noise contribution due to the amplifier, we have upgraded the readout chain by replacing these preamplifiers by the LARASIC4 \cite{DGeronimo} CMOS Integrated Circuits (ICs), which have been designed and realized for the MicroBooNE LAr TPC~\cite{Chen}.
The main features of these ICs are:
\begin{itemize}
\setlength{\itemsep}{3pt}
\item{16 channels per IC;}
\item{Low noise charge amplifiers with high-order filters;}
\item{Programmable gain: 4.7, 7.8, 14 and 25 mV/fC;}
\item{Programmable filter peaking time: 0.5, 1.0, 2.0 and 3.0~$\mu$s;}
\item{Power dissipation of < 10~mW/channel;}
\item{Specific design for long operation at cryogenic temperatures (15-20 years, based on accelerated aging test \cite{Sli}).}
\end{itemize}
\vspace{0.3cm}

\begin{figure}[ht]
\centering	
\includegraphics[width=0.55\linewidth]{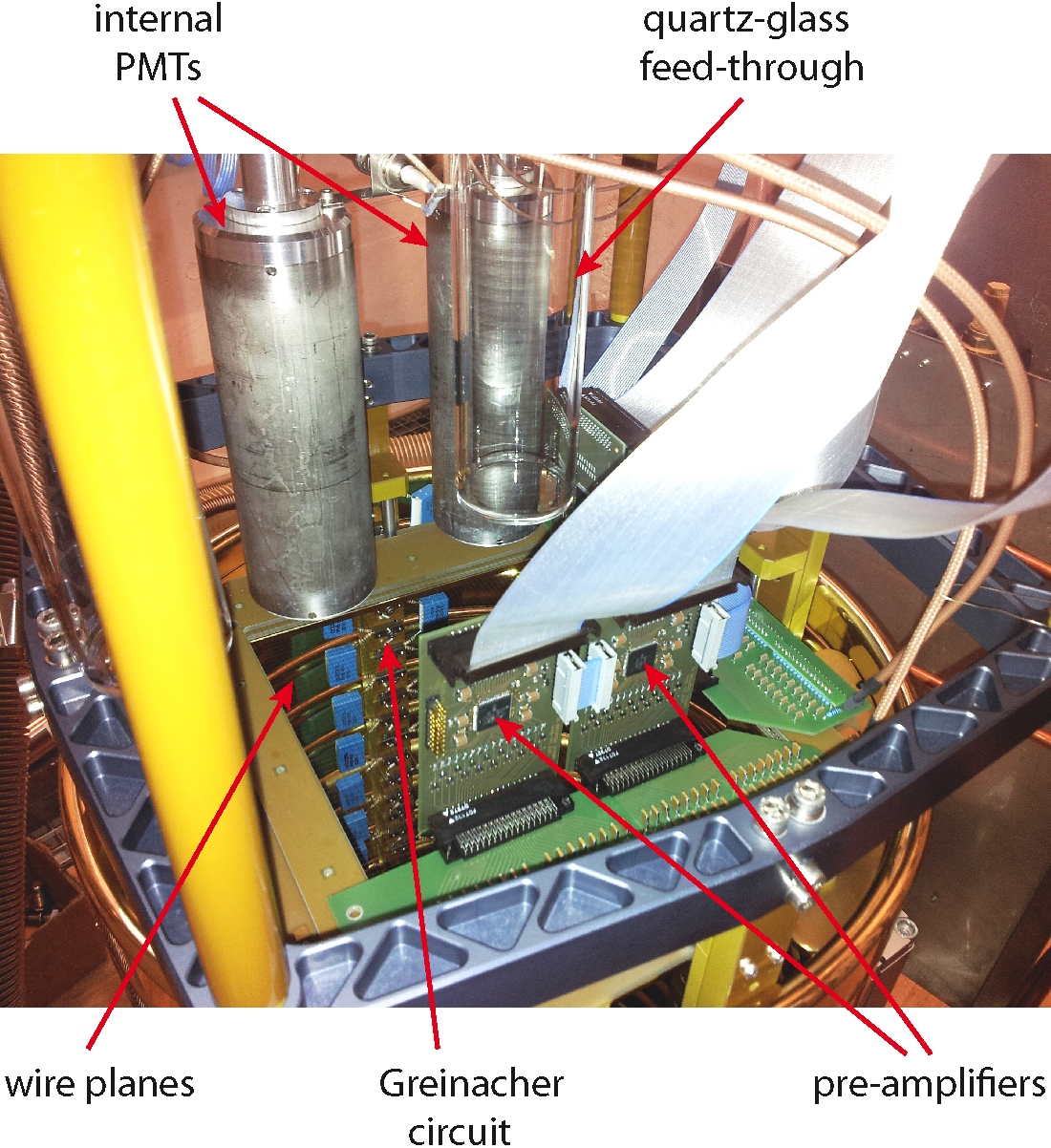}
\caption{LARASIC4 ICs mounted on the ARGONTUBE charge readout wire plane. Also shown are the two internal photo-multiplier tubes (PMT) of the light readout system, the Greinacher circuit used to generate the drift field in the TPC, and the optical feed-through of the laser calibration system (see~\cite{ARGONTUBE2} for details).}
\label{mount}
\end{figure}

In ARGONTUBE, the charge coming from the TPC wire readout plane (wire pitch of 3~mm, 64x64 channels) is  directly sent to the preamplifier PCBs, each hosting two LARASIC4 ICs. Four of these boards (8 ICs) provide charge amplification for a total number of 128 readout channels (see Figure~\ref{mount}). 

\begin{figure}[ht]
\centering	
\includegraphics[width=0.99\linewidth]{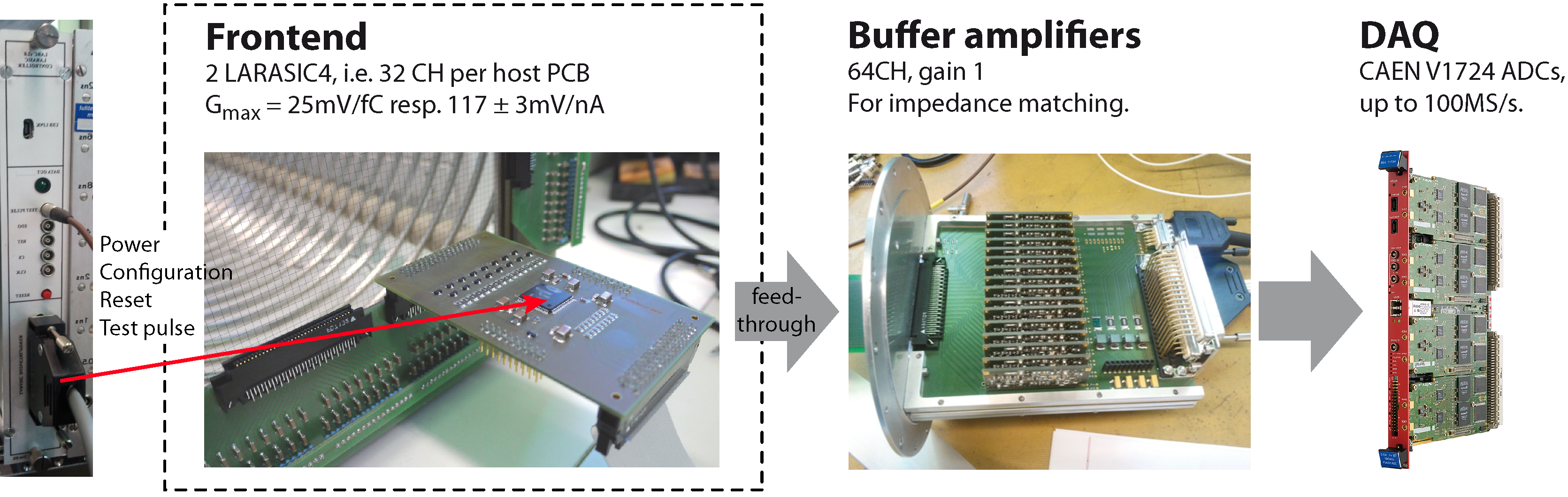}
\caption{Scheme of the ARGONTUBE charge readout chain. From left to right: digital control unit, preamplifier board at the readout wire plane, buffer amplifiers board, CAEN V1724 digitizer boards. The dashed rectangle denotes cold part of the chain.}
\label{chain}
\end{figure}

The full charge readout chain is shown in Figure~\ref{chain}. The leftmost panel is a picture of the 
NIM controller module used for programming the configuration registers of the LARASIC4 ICs and 
for providing power to the preamplifier boards. 
This module also sends the programmable calibration test pulse to the corresponding inputs of the 
preamplifier ICs. In the second panel, one preamplifier PCB is shown mounted on the wire charge 
readout plane. This part operates in the TPC at the temperature of liquid argon (87~K). The third 
panel shows the unit containing 64 buffer current amplifiers with a voltage gain of 1. Two of these units are mounted at the top flange of the cryostat and provide impedance matching between the LARASIC4 and the 50~$\Omega$ coaxial transmission lines, which deliver the analog signals to the digitizer boards (rightmost panel).

\begin{figure}[ht]
\centering	
\includegraphics[width=0.4\linewidth]{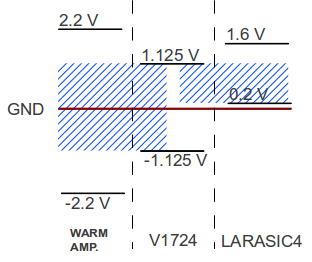}
\caption{Signal voltage mapping of outputs from warm hybrid (left) and LARASIC4 (right) preamplifiers 
to the input of CAEN V1724 digitizer board (center). Hashed zones indicate useful signal ranges. 
LARASIC4 output signal has rail-to-rail (0~V to 1.8~V) output swing, but the linearity is guaranteed only for 0.2~V to 1.6~V range. }
\label{vmap}
\end{figure}

The ASICs were configured with a charge sensitivity setting of G~=~25~mV/fC and a peaking time of T$_p$~=~3~$\mu$s. The trans-impedance in this configuration is measured to be 117$\pm$3~mV/nA for a constant current input signal. The sampling period of the digitizer was set to 1.01~$\mu$s/sample. 
The voltage mapping of outputs of both preamplifiers to V1724 digitizer input is shown in Figure \ref{vmap}. The nominal equivalent noise charge (ENC) of the ASIC in this configuration is equal to 375~$e^-$ in the absence of detector capacitance (at 77~K) according to the datasheet. Assuming a detector capacitance of 150~pF, the nominal ENC amounts to 1200~$e^-$ at 300~K and to 600~$e^-$ at 77~K.

\section{Detector performance}

\begin{figure}[ht]
\centering	
\includegraphics[width=1\linewidth]{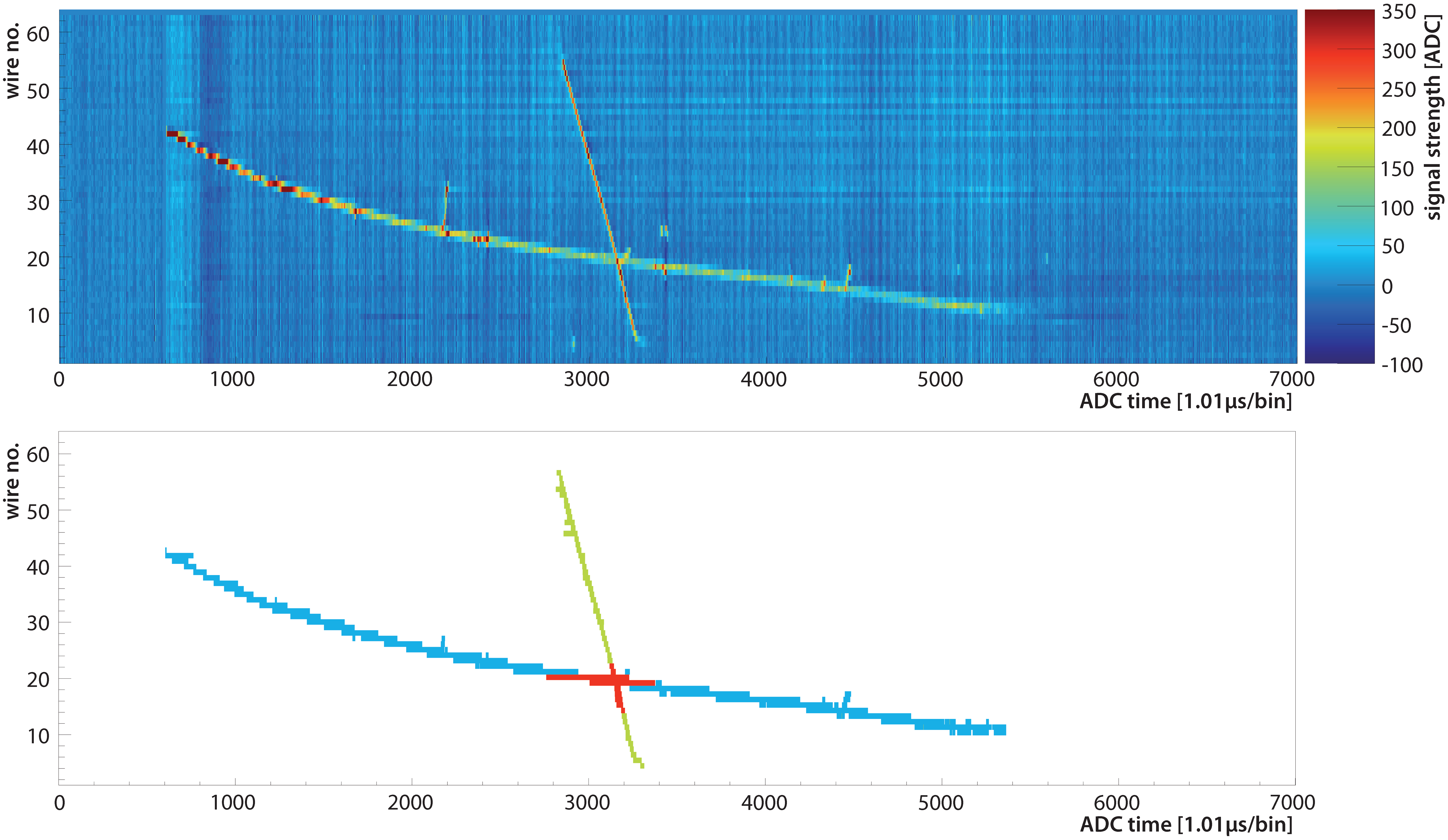}
\caption{Typical cosmic ray muon tracks reconstructed from ARGONTUBE data. Top: Collection wire plane raw data; Bottom: two reconstructed muon tracks. The red zone denotes ambiguous hits that may belong to either of the two tracks.}
\label{muon}
\end{figure}

\begin{figure}[ht]
\centering	
\includegraphics[width=1.\linewidth]{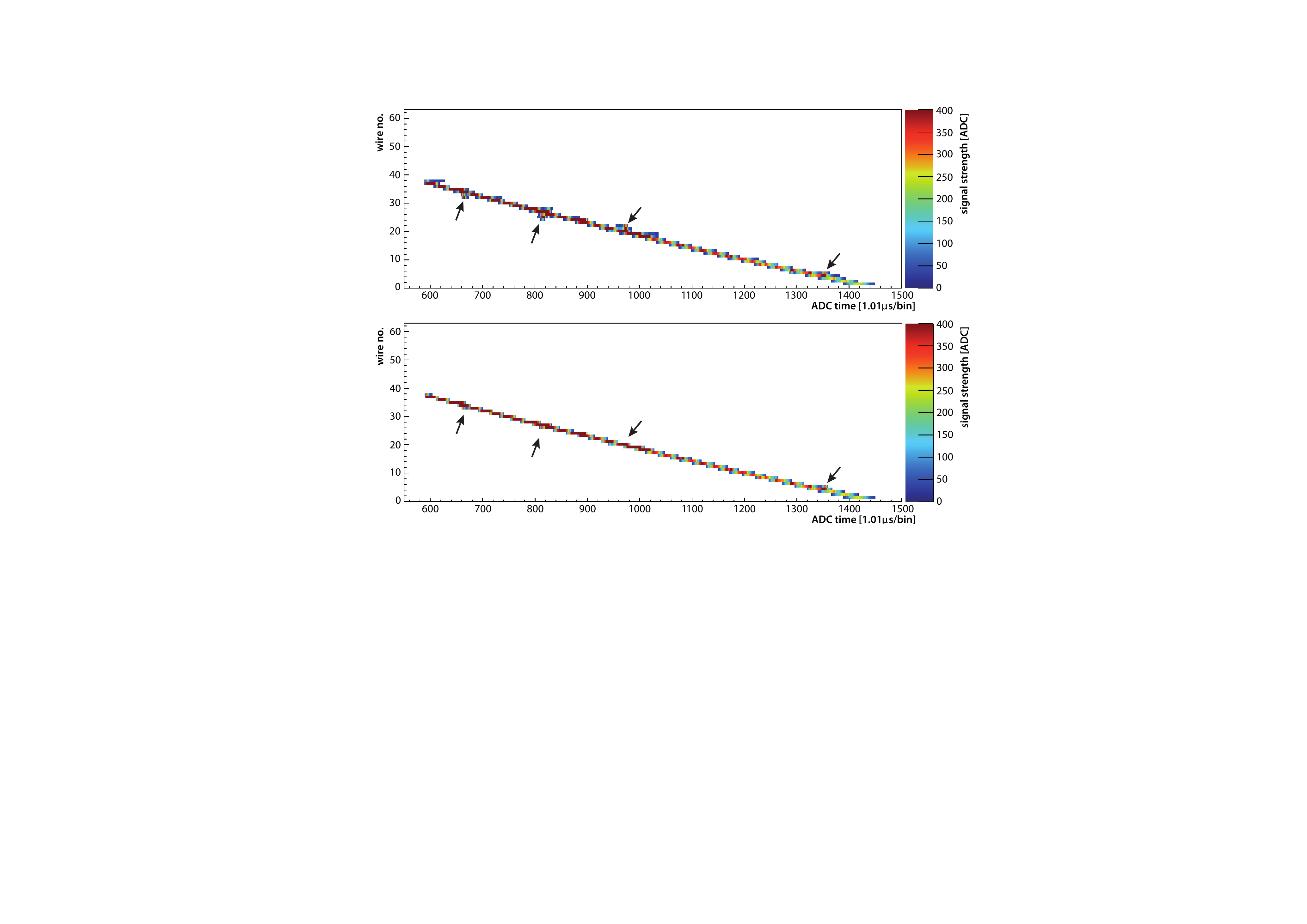}
\caption{Reconstructed high-momentum muon track with delta-electrons (top) and after their rejection (bottom).}
\label{deltas}
\end{figure}

More than 60000 cosmic muon events were recorded by the ARGONTUBE DAQ during a run performed in 2013. To 
cross-check the gains and characterize the readout performance, a dedicated track 
reconstruction algorithm was developed and successfully used. This algorithm is based on an 
optimized Hough-transform (PCLines algorithm, see~\cite{pclines} for details) implemented in 
the CUDA C language and run on a massively parallel NVIDIA 
GPU architecture\footnote{\url{www.nvidia.com/object/cuda_home_new.html}}. 
A typical reconstructed cosmic muon event is shown in Figure~\ref{muon}. Raw data from the 
collection wire plane with subtracted pedestals are shown in the upper plot, while the bottom one shows two identified,
overlapping muon tracks in the event. The red zone denotes ambiguous hits, which may belong to 
either track of the two. Only hits unambiguously assigned to a single track are used for further analysis. 
The next step of the reconstruction procedure is to free the track from associated delta-electrons. 
This is done by rejecting pixels located outside a narrow road (about 4$\sigma$) along the fitted straight track.  
Figure~\ref{deltas} shows a track before (top) and after (bottom) delta-ray rejection. Delta-electrons 
are indicated by black arrows. The data set is reduced to only straight tracks from high energetic muons with negligible scattering with respect to the coordinate resolution of the detector.

\begin{figure}[ht]
\centering	
\includegraphics[width=1.\linewidth]{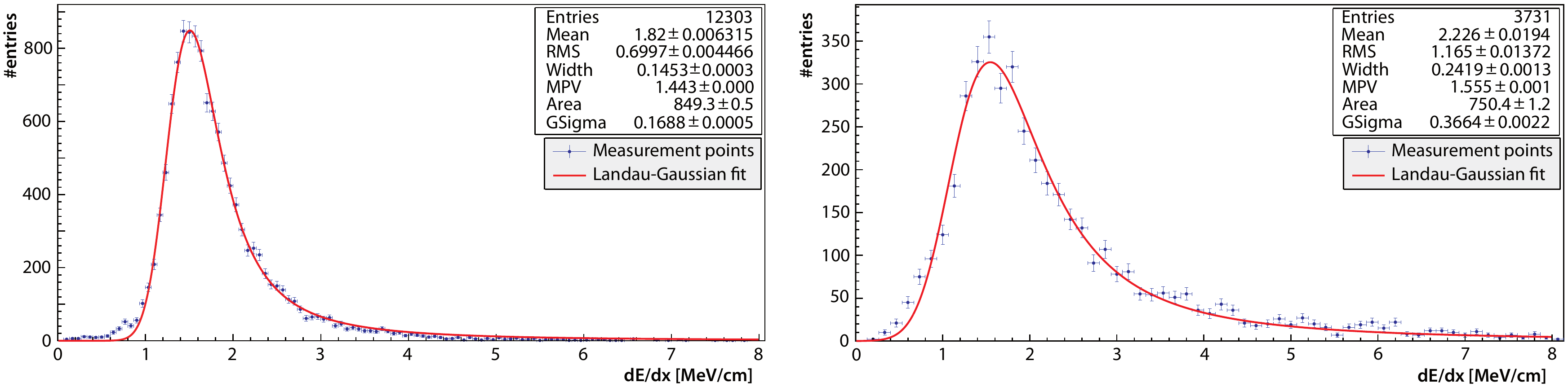}
\caption{Measured distributions of MIP tracks LET in ARGONTUBE, fitted with Landau-Gaussian function. Left: events recorded with the cryogenic LARASIC4 readout; Right: for comparison, a similar distribution obtained from data recorded with hybrid charge preamplifiers working at room temperature. The difference in width due to S/N ratio is well visible.}
\label{landau}
\end{figure}

To reconstruct particle tracks in 3D, data from the first (collection) and second (induction) 
projection wire planes have to be combined and processed appropriately. However, rather than matching 
every hit found in the collection plane to the corresponding one in the induction plane, a 
simplified approach is adopted, where the two projection angles of the track are determined from 
straight line fits to the hits of the collection and induction views separately. Once the track is 
reconstructed and freed from delta-electrons, its amplitude is corrected by the projection angles to 
build a Landau distribution of the linear energy transfer (LET) dE/dx of the ionizing particle. In the 
case of a MIP, the LET can be derived from the produced ionization charge \(dE/dx=dQ/dx\cdot\frac{W}{e~R}\), where $W$~=~23.6~eV is the energy needed to produce one electron-ion pair in liquid argon~\cite{wvalue}, $e$ is the electron charge, and $R$~=~0.35$\pm$0.04 (at 200~V/cm) is the drift field dependent charge recombination factor ~\cite{recomb}. The resulting distributions are shown in Figure~\ref{landau}. The left plot shows the results obtained
by using the cryogenic LARASIC4 charge preamplifiers. The one on the right shows for comparison the data obtained using the room temperature operated hybrid preamplifiers. The mean values for LET are 1.82$\pm$0.01(stat.)$\pm$0.34(sys.)~MeV/cm and 2.23$\pm$0.02(stat.)$\pm$0.42(sys.)~MeV/cm, respectively.

The discrepancy of the measured <dE/dx> values with respect to the expected value of 2.1~MeV/cm \cite{PDG} and to measured by other experiments ($e.g.$~\cite{argoneut}) is well within the errors for both cases. The most probable dE/dx values, 1.44$\pm$0.26(sys.)~MeV/cm and 1.56$\pm$0.27(sys.)~MeV/cm (with negligible statistical errors), are also in good agreement with that reported in literature \cite{palamara}. 

The following variables contribute to the systematic error of LET:
\begin{itemize}
\setlength{\itemsep}{3pt}
\item{The uncertainty on the charge preamplifiers' sensitivity and trans-impedance is obtained from test bench measurements using calibrated injected charge.}
\item{The uncertainty on the free electrons' lifetime is obtained from the dedicated measurements with cosmic muons and ionization tracks induced by UV laser beam \cite{ARGONTUBE1}.}
\item{The uncertainty of the track length per wire, which is derived from the detector spatial resolution. The resolution is defined by fit errors from the linear fit for each muon track.}
\item{The error on drift field calibration is estimated conservatively to be 7~V/cm \cite{ATField}. This defines the error of field-dependent charge recombination factor $R$, together with experimental errors on its parametrization ~\cite{recomb}.}
\end{itemize}
\vspace{0.3cm}

In order to get rid of the influence of V1724 board digitization noise, we used an oscilloscope to measure the MIP signal. The trigger was configured to target horizontal muons crossing the detector within 0.5~m from the readout plane.
The value of the MIP signal, measured by averaging the output with an oscilloscope, is 33.0$\pm$7.9~mV. 
As expected, it lies between the values derived from the average <dE/dx> (37.4~mV) and the most probable dE/dx (27.7~mV), and is compatible with both of them within the errors. This agreement confirms the validity of the gain calibration data that we have used throughout this study. The noise RMS was measured with the oscilloscope at the digitizer input in absence of a drift field in the TPC. The charge response 
and noise parameters together with the gain calibration data are summarized in Table~\ref{table1}. The
detector noise RMS measured with the LARASIC4 is equivalent to the ENC of 525~$e^-$. This value is in good 
agreement with the nominal one expected for the LARASIC4, taking into account the ARGONTUBE sensing wire 
capacitance of 7-10~pF.

Resulting dynamic range of the setup with the warm hybrid preamplifier is 2270. The dynamic range of setup with LARASIC4 depends on the configured gain of the IC and ranges from 440 at 25~mV/fC to 2340 at 4.7~mV/fC.

\begin{figure}[H]
\centering	
\vspace{0.4cm}
\includegraphics[width=1\linewidth]{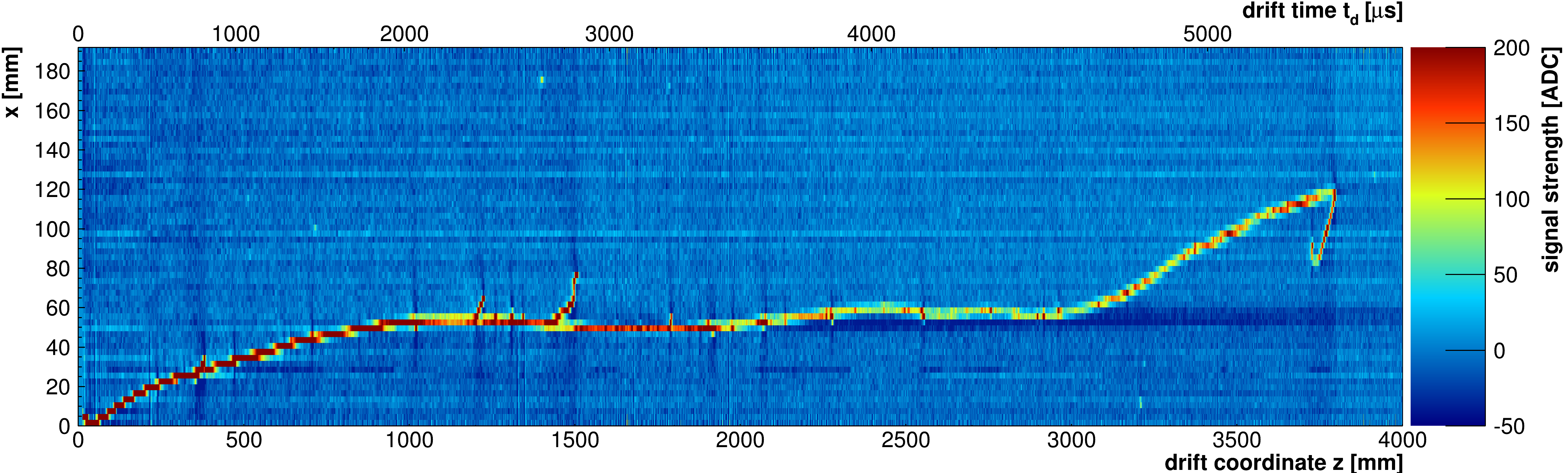}
\vspace{0.4cm}
\includegraphics[width=1\linewidth]{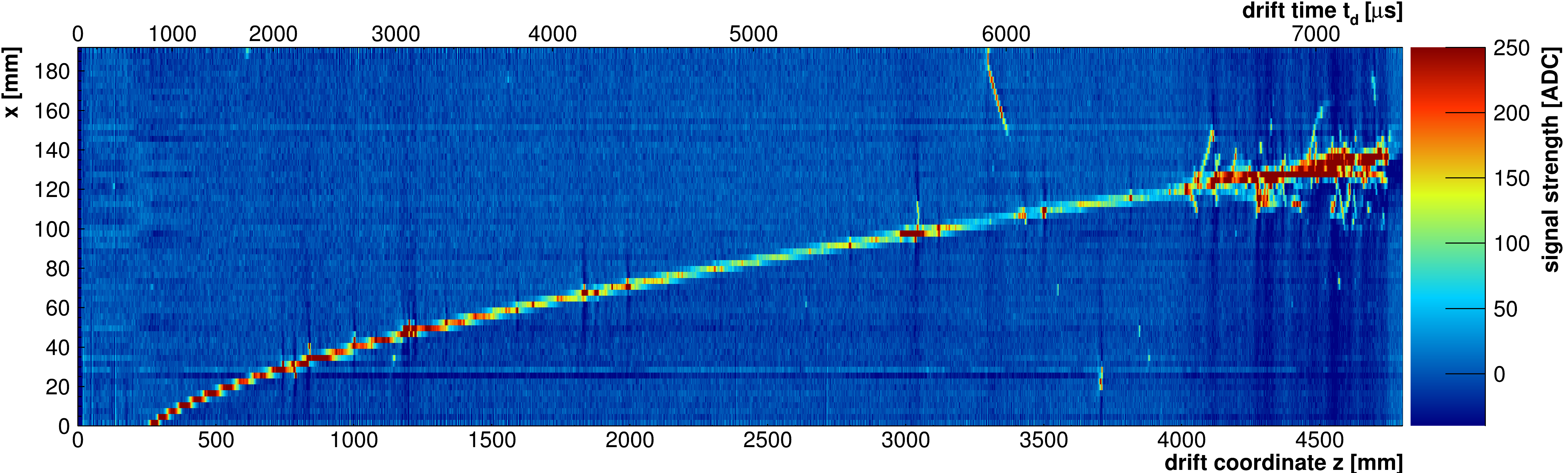}
\vspace{0.4cm}
\includegraphics[width=1\linewidth]{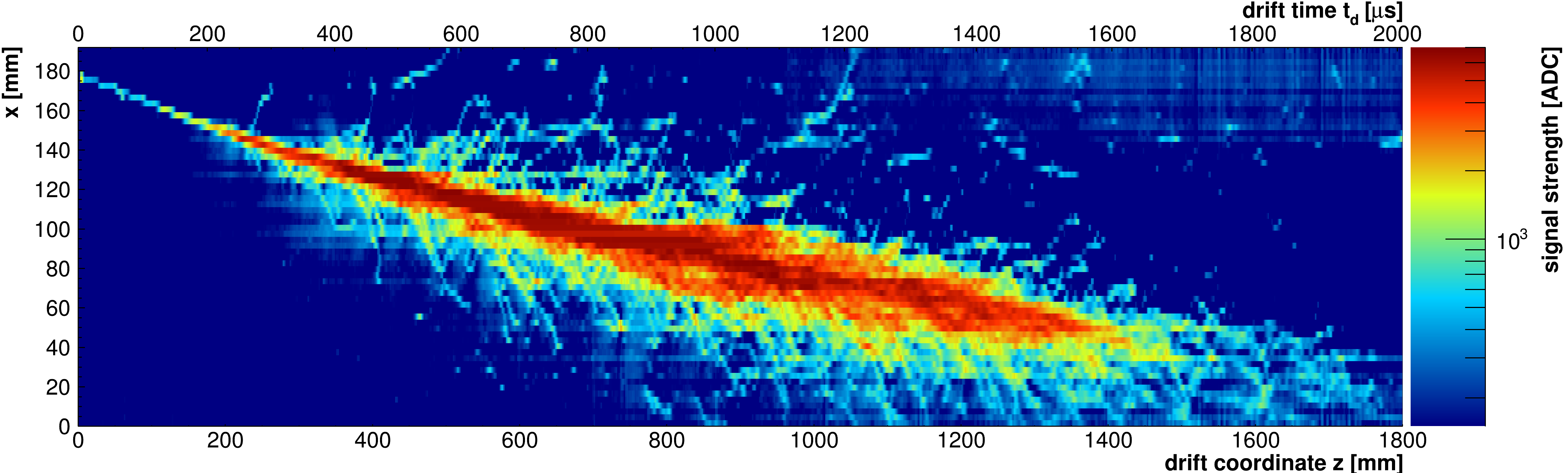}
\vspace{0.4cm}
\includegraphics[width=1\linewidth]{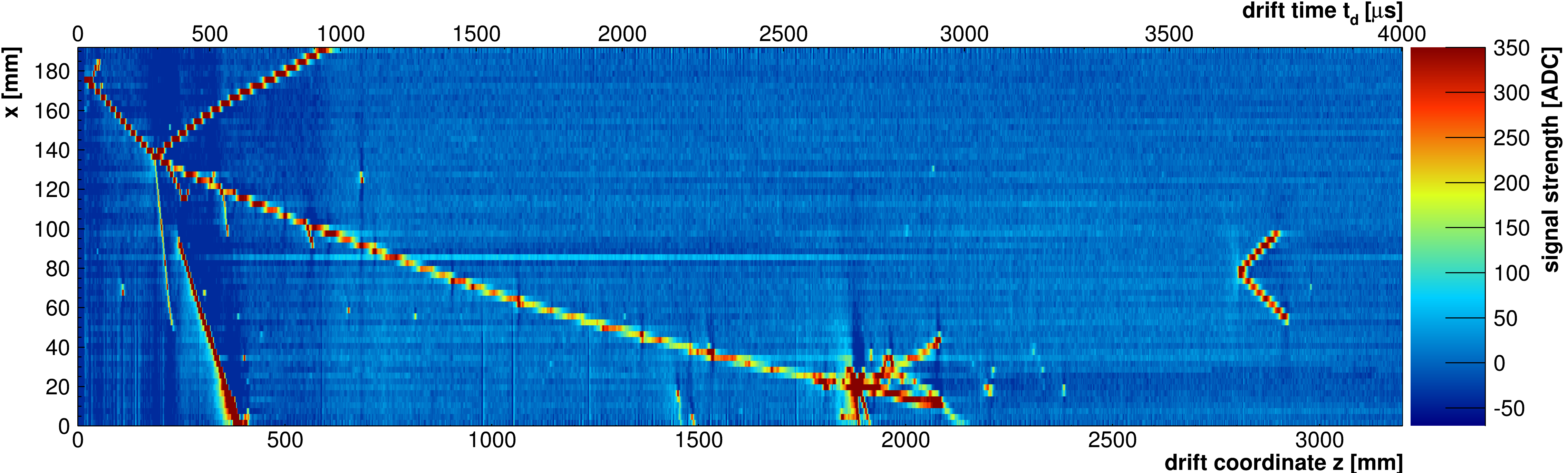}
\vspace{0.2cm}
\caption{Several cosmic ray events recorded with ARGONTUBE using LARASIC4 cryogenic charge preamplifier ICs: A muon stopping and decaying (top), a muon passing through the entire sensitive volume longitudinally and inducing an electromagnetic shower near the detector cathode (2nd from the top), an almost fully contained electromagnetic shower (3rd from the top) and a hadronic interaction (bottom).}
\label{events}
\end{figure}

For a drift field of about 200~V/cm, the signal-to-noise (S/N) ratio for the tracks induced by a MIP passing through the detector sensitive volume close to the wire planes is equal to 15.7$\pm$3.8 for the cryogenic charge preamplifiers. For the tracks at the far end of the detector, S/N ratio falls to about 2, corresponding to a free electron life time of about 3~ms. This allows to register and reconstruct particle tracks in the full 5~m long drift volume of the ARGONTUBE. A few examples of ARGONTUBE cosmic ray induced events are shown in Figure~\ref{events}.

\begin{table}[ht]
\caption{Summary of the charge calibration and noise analysis of the two different ARGONTUBE detector readouts obtained with cosmic ray muon data.}
\vspace{1mm}
\centering	
\begin{tabular}{llrr}
\toprule
\bf{Parameter}                    &          & \bf{LARASIC4}  & \bf{Hybrid pre-amp.}  \\
                                  &          & (at 87~K)      & (at 290~K) \\ 
\midrule
Charge sensitivity                       & [mV/fC]  & 25             & 2.1 \\ 
Trans-impedance                    & [mV/nA]  & 117            & 13 \\ 
Average MIP signal (oscilloscope) & [mV]     & 33.0$\pm$7.9   & 2.8$\pm$0.6 \\ 
Output noise (RMS)                       & [mV]     &  2.1$\pm$0.1   & 1.1$\pm$0.1 \\ 
Digitizer LSB equivalent                      & [mV]     &  0.15   & 0.15 \\ 
S/N ratio for MIP                         &          &  15.7$\pm$3.8  & 2.6$\pm$0.6  \\ 
Dynamic range selection               &          &  440 to 2340  & 2270  \\ 
Dynamic range (ARGONTUBE)  &          &  440  & 2270  \\ 
\bottomrule
\end{tabular}
\label{table1}
\end{table}

\section{Conclusions}
A factor of 6 improvement in signal-to-noise (S/N) ratio for tracks induced by minimum ionizing particles (MIP) was achieved by replacing hybrid charge preamplifiers operating at room temperature with dedicated CMOS ICs (LARASIC4) operating in liquid argon at 87~K and mounted directly on the wire charge readout boards of the ARGONTUBE TPC. For the cryogenic preamplifiers, the S/N ratio reached for particles crossing the detector in proximity of the readout planes was measured to be S/N~=~15.7$\pm$3.8 at a drift field of about 200~V/cm.

The equivalent noise charge (ENC) obtained in the configuration of an amplifier charge sensitivity of 25~mV/fC and a peaking time setting of 3~$\mu$s is measured to be 525~$e^-$. With either type of preamplifier, the mean and most probable values for the dE/dx of MIP in liquid argon obtained from data acquired by ARGONTUBE are both compatible with the literature. 

\section{Acknowledgments}
We warmly acknowledge the colleagues from the Brookhaven National Laboratory, V.~Radeka, H.~Chen, G.~De~Geronimo and S. Li, designers and developers of LARASIC4. They not only provided us with the required number of ICs for our tests, but also contributed to setting them in operation on the ARGONTUBE TPC.
We also thank C.~Bromberg and D.~Shooltz from Michigan State University for their kind help in designing the interface boards for LARASIC4.
We also express our gratitude to engineering and technical staff in Bern for the design, manufacturing and assembly of the detector.


\end{document}